\documentclass[amsmath,amssymb,twocolumn,floatfix,aps,showpacs,superscriptaddress,notitlepage]{revtex4-2}

\usepackage{amsmath,amsthm,amssymb,bm,hyperref,graphicx,color,float,changes,mathtools,enumitem}
\allowdisplaybreaks

\newcommand{\be}{\begin{equation}}
\newcommand{\ee}{\end{equation}}
\newcommand{\bea}{\begin{eqnarray}}
\newcommand{\eea}{\end{eqnarray}}
\newcommand{\6}{\partial}

\newcommand{\nuu}{N_\uparrow}
\newcommand{\ndd}{N_\downarrow}

\usepackage{pdfpages} 
\makeatletter
\AtBeginDocument{\let\LS@rot\@undefined}
\makeatother

\begin{document}

\title{Temperature-dependent periodicity of the persistent current in  strongly interacting
 systems}

\author{Ovidiu I. P\^{a}\c{t}u}
\affiliation{Institute for Space Sciences, Bucharest-M\u{a}gurele, R 077125, Romania}

\author{Dmitri V. Averin}
\affiliation{Department of Physics and Astronomy, Stony Brook University, Stony Brook, New York 11794, USA}

\begin{abstract}

The persistent current in small isolated rings enclosing magnetic flux is the current circulating in
equilibrium in the absence of an external excitation. While initially studied in superconducting and
normal metals, recently, atomic persistent currents have been generated in ultracold gases spurring a
new wave of theoretical investigations. Nevertheless, our understanding of the persistent currents in
interacting systems is far from complete, especially at finite temperatures. Here we consider the
fermionic one-dimensional Hubbard model and show that in the strong-interacting limit, the current can
change its flux period and sign (diamagnetic or paramagnetic) as a function of temperature, features
that cannot be explained within the single-particle or Luttinger liquid techniques. Also, the magnitude
of the current can counterintuitively increase with temperature, in addition to presenting different
rates of decay depending on the polarization of the system. Our work highlights the properties of the
strongly-interacting multi-component systems which are missed by conventional approximation techniques,
but can be important for the interpretation of experiments on persistent currents in ultracold gases.

\end{abstract}

\maketitle

\textbf{\textit{Introduction.}} The existence of  a persistent current (PC) in small metallic rings
threaded by magnetic flux $\phi$ at low temperatures has been theoretically predicted since the early
days of quantum mechanics \cite{Hun38} and superconductivity \cite{Blo65,Schi68,GY69,Ku70}. Following the
publication \cite{BIL83}, PCs were the focus of intense theoretical investigations, and were experimentally
confirmed in both invididual and ensembles of metal rings \cite{Lev90,Chan91,Jari01,Deb02,Blu09,Har09}.
Resurgent interest in the field is due to the generation of atomic PCs in ultracold
gases of single-component bosons \cite{Rye07,Wri13}, spinor bosons \cite{Bea13}, and
very recently of spinfull fermions \cite{Cai21}. Ultracold gases are characterized by an unprecedented
degree of control  over interaction strength, statistics, number of components and geometry,
allowing for the investigation of various properties of fundamental interacting models \cite{Chi15}.

The PC is a paradigmatic example of quantum coherence in mesoscopic systems and its magnitude is given
by $I(\phi)=-\6 F(\phi)/\6\phi$  with $F(\phi)=-k_B T \ln \mathcal {Z}(\phi)$ the free energy
and $\mathcal{Z}(\phi)$ the canonical partition function \cite{BY61,Blo70}. Gauge-invariance
implies periodicity, $I(\phi+\phi_0)=I(\phi)$, with $\phi_0$ the flux quantum $\phi_0=h/e$ ($h$ is the
Planck's constant and $e$ the charge of the electron) and from time-invariance we have $I(-\phi)=-I(\phi)$.
In addition to the amplitude of the current, defined by $I_{max}=\mbox{max}_{\phi\in (0,\phi_0/2)} |I(\phi)|$,
and periodicity, we are interested in the sign of the magnetic response: diamagnetic or paramagnetic. A
system is diamagnetic (paramagnetic) if $F(\phi)$ has a local minimum (maximum) at $\phi=0$.
While there is a large  body of work, mainly focused on free electrons with disorder, our understanding of
PCs in interacting systems, especially its temperature dependence, is far from complete. Using a variational
approach, Leggett  conjectured \cite{Lege91} that the ground-state energy of $N$ polarized interacting
fermions is diamagnetic for odd $N$ and paramagnetic for even $N$. This conjecture was proved and extended
to the case of small temperatures using Luttinger liquid (LL)  methods \cite{Loss92}.
At small temperatures and within the LL regime, the amplitude of the current decreases
exponentially with temperature, but the periodicity and sign of the current remain unchanged. For interacting
fermions with spin [$\ndd (\nuu)$ electrons have spin down (up)] a general result valid at arbitrary temperature
\cite{WFK08} is that $F(0)\le F(\phi/\phi_0)$ for $N_{\downarrow,\uparrow}$ both odd, and $F(1/2)\le F(\phi/\phi_0)$
for $N_{\downarrow,\uparrow}$ both even. This result does not preclude periodicities smaller than $\phi_0$ or
changes in the sign of the current with temperature, as we will show below.

\textbf{\textit{The Hubbard model in a magnetic field.}}
We consider a system of $N$ electrons of which $\ndd$ have spin down on a ring lattice with $L$ sites
and repulsive interactions. The ring is threaded by an Aharonov-Bohm  flux $\phi$. The system is described
by the Hubbard Hamiltonian \cite{LW, Zvya90, SS90, EFGKK}
\begin{align}\label{ham}
\mathcal{H}=-t\sum_{j=1}^L\left[\sum_{\sigma}\left(e^{-i eA}c_{j+1,\sigma}^\dagger c_{j,\sigma}
+\mbox{h.c.}\right)
 -\frac{U}{t} n_{j,\uparrow}n_{j,\downarrow}\right],
\end{align}
with $n_{j,\sigma}=c_{j,\sigma}^\dagger c_{j,\sigma}$ the number of electrons of spin $\sigma=\{\downarrow,\uparrow\}$ at site $j$.
In (\ref{ham}), $A=2\pi \phi/ (L\phi_0$) is the vector potential of the magnetic flux, $t$ is the electron hopping integral
and $U>0$ quantifies the strength of the repulsive interaction. In the following, we will measure the energies in units of $t$
and set $\hbar=1$ and $e=1$.
\begin{figure*}[th]
 \centering
 \includegraphics[width=1\linewidth]{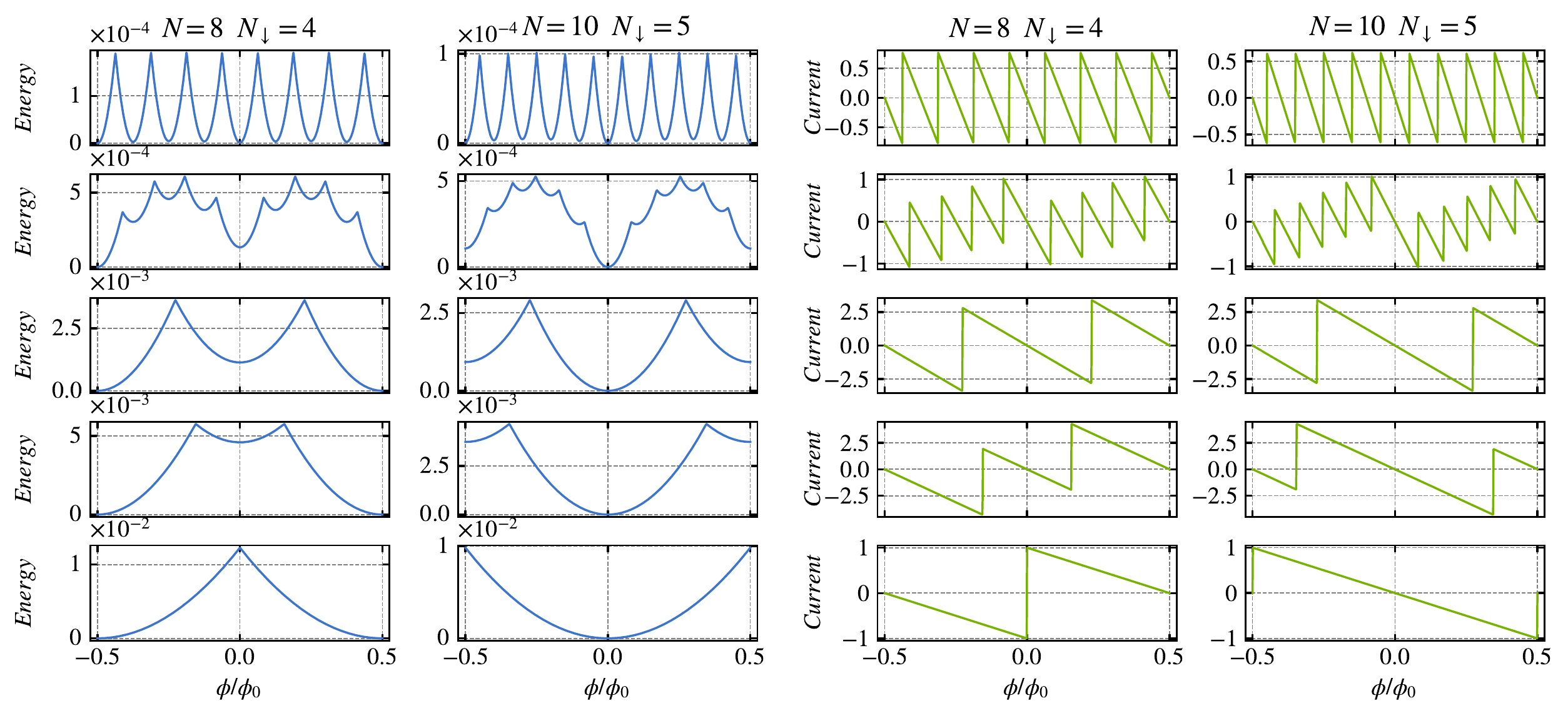}
 \caption{ Dependence of the ground-state energy (left columns; normalized
          by $t$) and current (right columns; in units of $I_0(\ndd)+I_0(\nuu)$) on the magnetic flux and strength of the interaction for $N=8, \ndd=4$ and $N=10, \ndd=5$. From top row to bottom the coupling strength is $U=5000, 50, 5.5, 1 ,0$ and the density is the same in both cases, $n=0.1$. }
 \label{figT0} \end{figure*}
The Hamiltonian (\ref{ham}) is exactly solvable with the Bethe ansatz equations (BAEs) \cite{SS90}:
$
k_jL=2\pi I_j+2\pi\phi/\phi_0-\sum_{\alpha=1}^{\ndd}\theta\left(\frac{\sin k_j-\lambda_\alpha}{u}\right)\, ,$
$
\sum_{j=1}^N\theta\left(\frac{\lambda_\alpha-\sin k_j}{u}\right)=2\pi J_\alpha+\sum_{\beta=1}^{\ndd}
\theta\left(\frac{\lambda_\alpha-\lambda_\beta}{2u}\right)\label{bael2}\,$
where $u\equiv U/4t$, $j=1,\cdots,N$, $\alpha=1,\cdots,\ndd$, $\theta(x)=2 \arctan(x)$ and $I_j=\ndd/2\,\  ( \mbox{mod } 1)$,
i.e.,  $I_j$ is integer or half-integer depending on whether $\ndd$ is even or odd, and similarly,
$J_\alpha=(N-\ndd+1)/2\,\ ( \mbox{mod } 1)$. The energy and momentum of a state are $E=-2\sum_{j=1}^N\cos k_j\, $ and
$P=\left[\sum_{j=1}^N k_j\right]\mbox{ mod } 2\pi .$ To find the PC we need to solve the BAEs
for the ground state (at $T=0$) or for all the relevant excited states (at $T>0$).
Needless to say, even the numerical investigation of the BAEs is in general very difficult. In
the most interesting case, the strong interaction limit, $U\gg 1,$ certain simplifications allow, however,
for a thorough investigation. We look first at some limiting cases.

\textbf{\textit{The $U=0$ case.} }
Because the particles are non-interacting, it is sufficient to consider spinless fermions. For fermions
with spin, the PC is then given simply by the sum of the contributions from the two spin directions \cite{LG91}.
The momenta of $M$ spinless fermions in the ring are $k_j=2\pi\left(j+\phi/\phi_0\right)/L,$ with $j$ integer.
This implies that at $\phi=0$, the ground-state is degenerate for even $M$, but not for odd  $M$, i.e., the
PC depends on the parity of the number of particles. In terms of the ``Fermi vector'' $k_F(M) \equiv \pi M/L$
and the Fermi velocity $v_F(M)= 2 \sin k_F(M)$, the PC of spinless fermions is \cite{CGRS88}: $I_{FF}(M,\phi)=-\frac{I_0(M)}{\sin\pi/L}\sin\left(\frac{2\pi}{L} \frac{\phi}{\phi_0}\right)\, $ for $M$ odd,
and $I_{FF}(M,\phi)=\frac{I_0(M)}{\sin\pi/L}\sin\left[\frac{\pi}{L}\left(1-\frac{2|\phi|}{\phi_0}\right)\right]
\mbox{sgn}\, \phi $ for $M$ even with $I_0(M)= ev_F(M)/L$. Therefore, the PC for electrons with spin is $I(\phi)=I_{FF}(\ndd,\phi)+I_{FF}(\nuu,\phi)$. When $N_{\downarrow,\uparrow}$ are both odd, $I(\phi)$ is
diamagnetic; when $N_{\downarrow,\uparrow}$ are both even, it is paramagnetic.

\textbf{\textit{The $U=\infty$ case.} }
In this ``impenetrable'' limit \cite{OS90,Kus91a,YF92,CHKA20}  the BAEs for the $k_j$'s become  $k_j^\infty=2\pi
\left(I_j+\phi/\phi_0+ \sum_{\alpha=1}^{\ndd} J_\alpha/N\right)/L\, ,$ which are equivalent to the result
for spinless fermions in a ring threaded by a magnetic flux $\phi/\phi_0+\sum_{\alpha=1}^{\ndd} J_\alpha/N.$
In the ground-state, $I_j$'s  fill an interval between $I_{min}$ and $I_{max}$, resulting in the energy
$ E=-2\frac{\sin(\pi N/L)}{\sin(\pi/L)}\cos\left[\frac{2\pi}{L}\left(\frac{\phi}{\phi_0}+\frac{1}{N}
\sum_{\alpha=1}^{\ndd} J_\alpha+D\right)\right]\, $
where $D\equiv (I_{min}+I_{max})/2$. This formula is valid for densities $n=N/L<1$ (at half-filling, $n=1$,
the sine factor in $E$ gives 0). The energy is minimized by choosing the set $\{J_\alpha\}_{\alpha=1}^{\ndd}$
such that $\sum_{\alpha=1}^{\ndd}J_\alpha=-p$ for $ (p-1/2)/N<\phi/\phi_0+D<(p+1/2)/N\, .$ This implies that
the PC for the not-fully-polarized impenetrable system at zero temperature is: a) periodic with a period of
$1/N$ of the flux quantum (this remains valid in the case of $SU(\kappa)$ fermions  with $\kappa> 2$ as
shown in a recent study \cite{CHKA20}),  b) diamagnetic, and c) does not present parity effects \cite{Kus91a,YF92} - see also
Fig.~\ref{figT0}. Note that for fully spin-polarized electrons, $\ndd=0$, the system is effectively non-interacting
even for large $U$, and PC is described by the same expressions as for $U=0$, e.g., has the period of one flux
quantum. This abrupt change of the flux period (from 1 to $1/N$) between the polarized electrons and a system with
even one flipped spin can be understood in terms of the change of the rotation period of the electron system in
real space \cite{AB18}. For polarized electrons, the rotation period is $1/N$ of the full rotation, while a
spin flip changes this period to a full rotation.

\textbf{\textit{First correction.}  }
For $u$ large but finite, the charge momenta with accuracy $1/u$ are \cite{YF92,KWKT94,EFGKK}
$k_j=k_j^\infty+\Delta k_j/u$ with $\Delta k_j=E_s\sum_{n=1}^N\left[\sin k_j^\infty-\sin k_n^\infty\right]/L$,
where $E_s=-2\sum_{\alpha=1}^{\ndd}1/[N(\Lambda^2_\alpha+1)]$ is the energy per lattice site of an
antiferromagnetic XXX Heisenberg spin-chain with spin rapidities $\{\Lambda_\alpha\}_{\alpha=1}^{\ndd}$
satisfying the BAEs: $N\theta(\Lambda_\alpha)=2\pi J_\alpha+ \sum_{\beta=1}^{\ndd}\theta[(\Lambda_\alpha
-\Lambda_\beta)/2]\, .$ The  energy with the same $1/u$ accuracy is
$
E=E^\infty+\frac{2E_s}{Lu}\left[N\sum_{j=1}^N\sin^2 k_j^\infty-\left(\sum_{j=1}^N\sin k_j^\infty\right)^2\right]
$
with $E^\infty=-2\sum_{j=1}^N\cos k_j^\infty$. Note that the energy depends on $\phi$ via $k_j^\infty$.
Thus, in the strong-coupling limit, the spin degrees of freedom are described by an anti-ferromagnetic Heisenberg
spin chain with the coupling constant $\sim \ndd/(Lu)$, while the charge degrees of freedom are similar to
free fermions. This regime, in which the energies  of the charge and spin sectors satisfy $E_{charge}\gg E_{spin}$,
is called the spin-incoherent regime \cite{BL87,B91,CZ1,CZ2,Matv,FB,F}. It is particular to multi-component systems,
and presents universal properties which are different from the LL case. For the Hubbard model,
the lowest energy states in this regime are obtained \cite{YF92} by considering the same distribution of $I_j$'s
as in the infinite repulsion case and finding the states of the Heisenberg chain with the lowest energy for a given
momentum $q=2\pi \sum_{\alpha=1}^M J_\alpha/N$, which are the des-Cloizeaux-Pearson excitations \cite{dCP62}.

The dependence of the ground-state energy and current on the strength of interaction at zero temperature is
shown in Fig.~\ref{figT0} for  ``balanced''  systems with $N=8$ and $N=10$. At very  strong coupling ($U=5000$) the
current is diamagnetic and has other characteristic features of the impenetrable case: $1/N$ periodicity and no parity
effects. For weaker interaction, the contribution of the spin sector becomes more pronounced, as manifested by the raising
of the $N$ parabolas. We can see that for $U=50$ and $U=5.5$, while the PC is still diamagnetic, the periodicity changes
to $1/2$.  At very weak interaction, the periodicity  becomes 1, and for $U=0$, one obtain a paramagnetic (diamagnetic)
current for $N_{\downarrow,\uparrow}$ even (odd). Therefore, the strength of the interaction has a strong influence
on the zero-temperature PC, changing its periodicity, amplitude, and even the sign (paramagnetic or diamagnetic).
Also, the $1/2$ periodicity seen in Fig.~\ref{figT0} at strong and intermediary coupling ($U=50$ and $U=5.5$) is a
particular case of the additional $\ndd/N$ periodicity characterizing the PC of the strongly interacting Hubbard model
first discovered in \cite{Kusm95} for $\ndd\ll N$. Figure \ref{figT0}  provides a proof of this additional $\ndd/N$
periodicity in the microscopic regime of small $N$. From the point of view of the real-space rotations, this periodicity
can be viewed as a manifestation of the antiferromagnetic order, which for the balanced electron system makes the rotation
period $1/N_{\downarrow}$ of the full rotation, i.e. two times the period $1/N$ for spinless electrons.
\begin{figure}[t]
 \centering
 \includegraphics[width=1\linewidth]{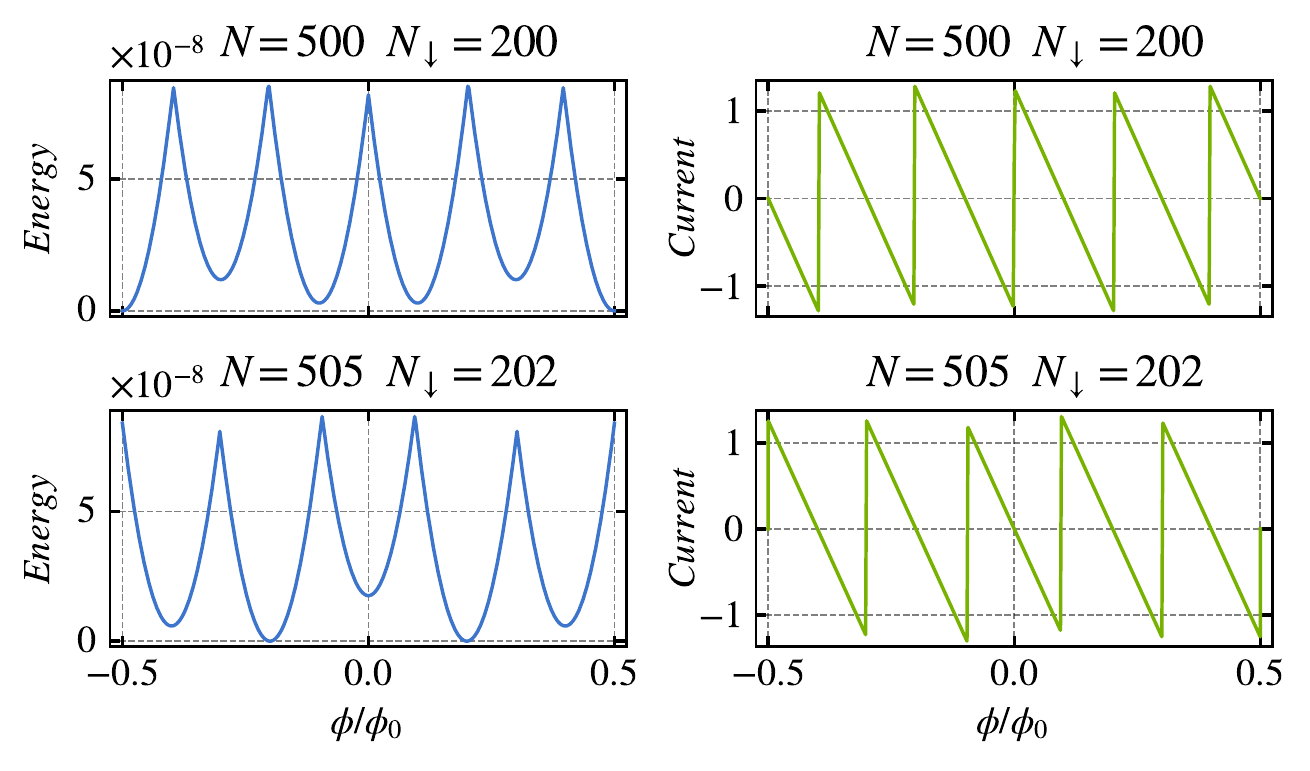}
 \caption{ Dependence of  energy (left column; normalized by $t$)
          and current (right column; in units of $I_0(\ndd)+I_0(\nuu)$)
          for mesososcopic systems with $N=500, \ndd=200$ and $N=505, \ndd=202$.  The strength of the interaction and density
          are $U=100$ and $n=0.01$. Note the change in sign of the PC. }
 \label{figMes}
\end{figure}

In the mesoscopic regime ($N,L\gg 1$), and  at zero temperature, the calculation of the persistent current is equivalent to the calculation
of the finite size corrections to the energy due to the change of the boundary conditions from periodic to twisted \cite{YF92}.
 For any value of $U$  and $n<1$, and assuming that
the $I_j$'s and $J_\alpha$'s are consecutive numbers, the corrections to the ground-state energy due to the magnetic flux are \cite{Woyn89,YF92,Zvya05}
\begin{figure*}[t]
 \centering
 \includegraphics[width=1\linewidth]{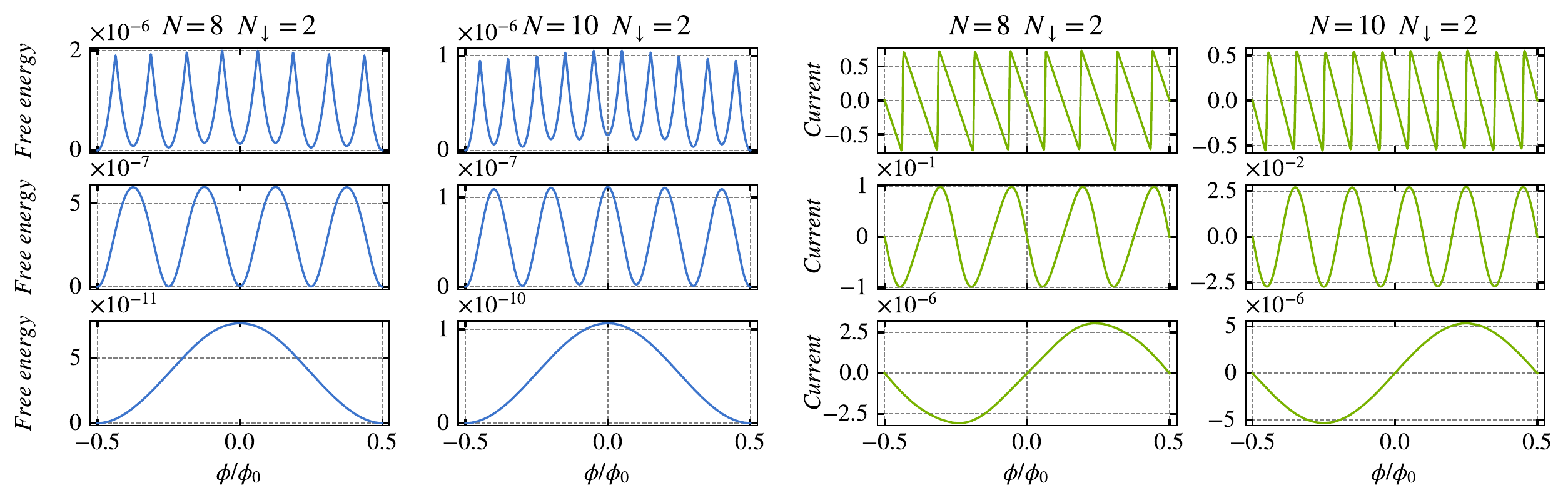}
 \caption{ Dependence of the free energy (left columns in units of $t$) and current (right columns normalized by $I_0(\ndd)+I_0(\nuu)$)
          on the magnetic flux and temperature for $N=8, \ndd=2$ and $N=10, \ndd=2$.
          From top row to bottom the temperatures are $T/T_s=1.01\, ; 48.1\, ; 582\, $ for $N=8$ and $T/ T_s=1.01\, ; 48.1\, ; 455\,$
          for $N=10$. The density and interaction strength are $n=0.01$ and $U=100$. }
 \label{figT}
\end{figure*}
\begin{align}\label{currentT0}
\Delta E(\phi)=&\frac{2\pi v_c}{L}\left[Z_{cc}\left(D_c+\frac{\phi}{\phi_0}\right)+Z_{sc}D_s\right]^2\nonumber\\
&\ \ +\frac{2\pi v_s}{L}\left[Z_{cs}\left(D_c+\frac{\phi}{\phi_0}\right)+Z_{ss}D_s\right]^2\, ,
\end{align}
where $Z_{cc},Z_{cs},Z_{sc},Z_{ss}$ are the matrix elements of the dressed charge matrix (see \cite{SM}), $v_{c,s}$ are the
velocities of the charge and spin excitations and $D_c$ and $D_s$ satisfy the constraints
$I_{max}-I_{min}+1=N\, ,$ $I_{max}+I_{min}=2D_c\, ,$ $J_{max}-J_{min}+1=\ndd\, ,$ $J_{max}+J_{min}=2D_s$.

When $U\gg 1$ we have $Z_{cc}\sim 1$, $Z_{cs}\sim 0$, $Z_{sc}\sim \ndd/N$ and $Z_{ss}$ is given by a simple integral
equation \cite{FK90,FK91}. In the same limit the  charge and spin velocities behave like $v_c\sim 2\sin \pi n,$ $ v_s\sim 1/U,$
which means that in the first approximation we can neglect the second term of (\ref{currentT0})
\be\label{currentT0H0}
\Delta E(\phi)=\frac{2\pi v_c}{L}\left[\left(D_c+\frac{\phi}{\phi_0}\right)+\frac{\ndd}{N}D_s\right]^2
\, .
\ee
When $\ndd/N=1/m$ with $m=2,3,\cdots$, this expression shows that the current has a $\ndd/N$ periodicity.
When $\ndd/N$ is not very close to $1/m$, the situation is more complicated (take into account that $D_c$ and $D_s$ have
integer or half-odd integer values depending on the parities of $\ndd$  and $N$).  For example, when $\ndd/N=2/5=0.4$  one would expect $1/2$ quasi-periodicity
(see \cite{Kusm95}), but in fact we obtain periodicity $1/5$ as can be seen in
Fig.~\ref{figMes}. Note also that the sign of the current depends on the parities of the number of particles: when $\ndd/N=2/5$
with $N,\ndd$ both even we have paramagnetic current while for $N$ odd and $\ndd$ even the current is diamagnetic. The same
features are present in microscopic systems with $N=10, \ndd=4$ and $N=5, \ndd=2$.

\textbf{\textit{Persistent current at finite temperature.}}
Computing thermodynamics of the Hubbard model, even in the thermodynamic limit, is a very difficult task, and it is
sensible to assume that computing the $1/L$ corrections is outside the reach of analytical methods. In the strong
coupling limit, however, one can take advantage of the fact that the energy of the spin sector is much smaller than
the energy of the charge sector, which allows for direct computation of the canonical partition function at low
temperatures by summing over all the spin eigenstates  and only some of the charge excitations. For a dilute system
($n<0.1$) the relevant temperature scales are $T_F=\pi^2 n^2$  for the charge degrees of freedom and $T_s=\pi^2 n^3/U$
for the spin degrees of freedom. For temperatures $T\ll T_F$ the partition function can be computed as
$
\mathcal{Z}(\phi)=\sum_{ \mbox{relevant sets } I}\;  \sum_{ \mbox{all sets } J} \exp \{- E(\{ k_j\},\phi)/T\}\,
$
and gives the PC. This approach requires the knowledge of the all $C^N_{\ndd}$ states of the Heisenberg spin-chain
with $N$ sites and $\ndd$ spins down, which can be found in \cite{KM97,HC07,HNS13}. Using this method we were able
to investigate the PC for all systems with $N\le 10\, , \ndd\le N/2$ and $T<0.06\, T_F$. While below we focus on
dilute systems, we note that our results remain valid for all densities $0<n<1$, if $U/n\gg 1$ (see \cite{SM}).

The dependence of the PC on temperature in the strongly interacting Hubbard model is very complex with the polarization
of the system playing an important role.  For a system with $N=8$ and $\ndd=2$, Fig.~\ref{figT} shows that while at very
low temperatures the current is diamagnetic with period $1/8$, at higher temperatures the periodicity changes to $1/4$. At
even higher temperatures the current becomes paramagnetic with period $1$. The evolution of the current with increasing
temperature for a system with $N=10$ and $\ndd=2$ is similar (but note paramagnetic current at  intermediate temperatures):
diamagnetic with period $1/10$, paramagnetic with period $1/5$, and paramagnetic with period $1$. Therefore,
we see the following pattern: in the ground-state, the current is diamagnetic with periodicity $1/N$, and is followed
at higher temperatures by the current with $\ndd/N$ periodicity whose sign is the same as the one at zero temperature,
which can be derived from (\ref{currentT0}). At very high temperatures, the current should have the same characteristics
as for free fermions with spin, i.e., for both $N_{\downarrow,\uparrow}$ even, paramagnetic with period $1$. This general
pattern can be understood by noting that an increase in temperature is qualitatively similar to the decrease in $U$, and
therefore the evolution of the current with $T$ mimics the evolution of the current at $T=0$, when interaction decreases.
For instance, the doubling of the current period from $1/N$ to $\ndd/N$ in Fig.~\ref{figT} can again be related to the
change of the rotation symmetry of the electron system in real space, from full rotation at low temperatures, to half of
the rotation at the intermediate temperatures, when the two spin-down electrons become located symmetrically in the system.

\begin{figure}
 \centering
 \includegraphics[width=1\linewidth]{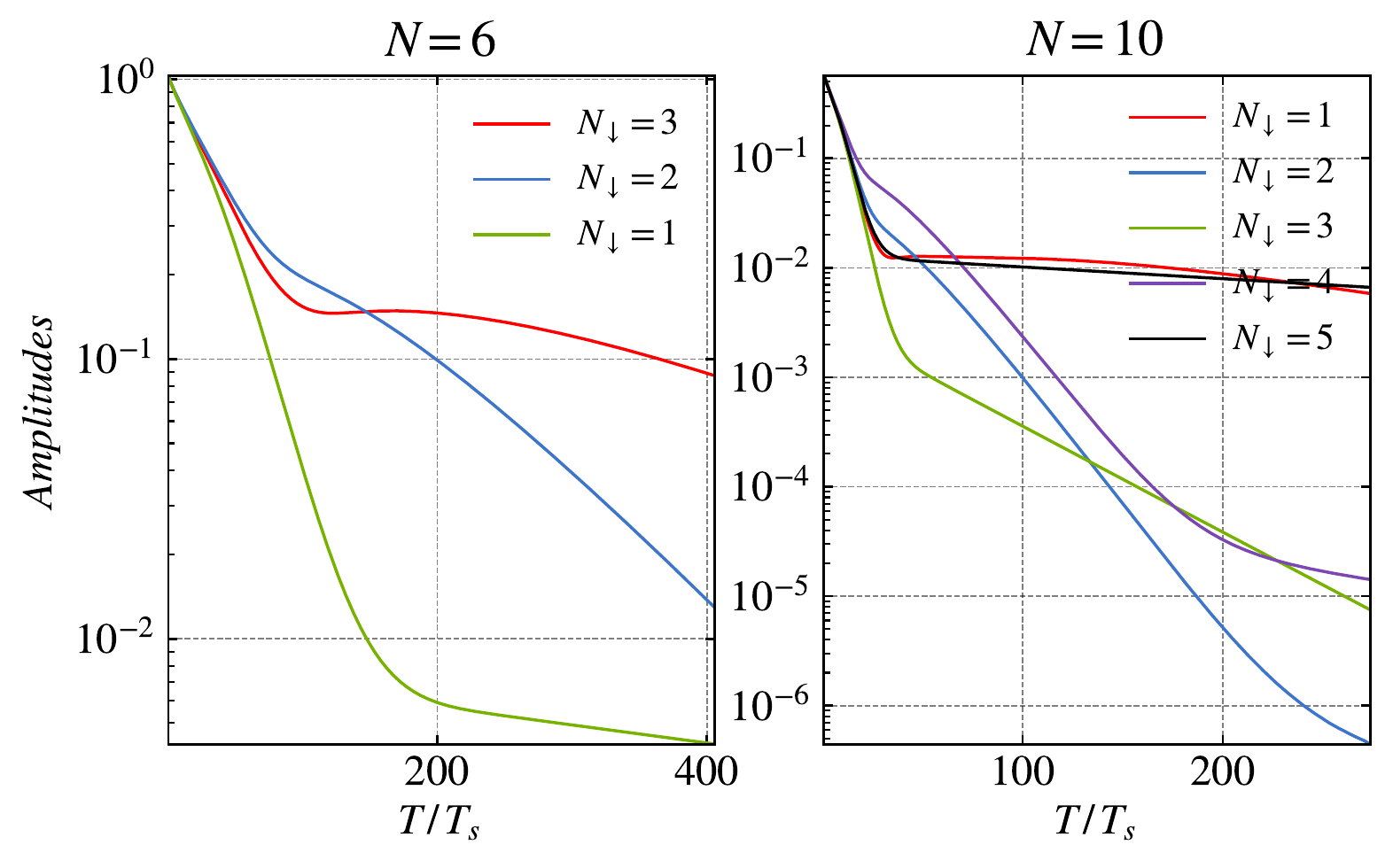}
 \caption{ Temperature dependence of the amplitudes (normalized by $I_0(\ndd)+I_0(\nuu)$) for $N=6$  and $N=10$. Note that
 for $N=6, \ndd=3$, and $N=10, \ndd=5$, there is an interval in which the amplitude is increasing with temperature. For
 all cases $n=0.01$ and $U=100$.}
 \label{figAmp}
\end{figure}

Qualitatively, the amplitude of the PC is reduced exponentially with increasing temperature. The quantitative temperature
dependence of the amplitude is plotted in Fig.~\ref{figAmp} for systems with $N=6$ and $N=10$, and shows that there
are different rates of decay associated with the different ranges of the system parameters: Luttinger liquid, spin-incoherent,
and almost free. The fastest rate of decay is in the LL regime; in the spin-incoherent regime, the rate of decay
depends strongly on the polarization of the system. A very interesting feature which can be seen in Fig.~\ref{figAmp}
is the presence of an interval of temperature in which the amplitude is {\em increasing} with $T$ for
$N_{\downarrow,\uparrow}=3$ and $N_{\downarrow,\uparrow}=5$. This counterintuitive feature is present at the transition
between the LL and the spin-incoherent regime and is due to the fact that the tail of the momentum distribution gets
strongly suppressed as the temperature increases (for a similar phenomenon  in spinless fermionic systems with non-trivial
geometry or dissipation see \cite{MoskD00,Mosk00}). This momentum reconstruction was first noticed in the case of the
Gaudin-Yang model (obtained in the dilute limit of the Hubbard model) in \cite{CSZ05}.  In details, the PC is on the
order of that produced by an electron at the Fermi level, $I=e v_F/L$. At finite temperature, the occupation
probabilities of levels close in energy to the Fermi level, and producing positive and negative contributions to the
current, are not very different, leading to suppression of the current associated with the broadening of the momentum
distribution. At the LL-spin-incoherent transition, however, the fraction of particles with higher momenta decreases,
resulting in a softer rate of decay for some polarizations, or even increase in the magnitude of the PC.

We expect that the temperature dependent periodicity and the different rates of decay of the PC to be general features
of strongly interacting fermionic systems with spin-independent interactions. The reason is that such systems
(integrable and non-integrable) present both the LL and spin-incoherent regimes, as it can be seen more explicitly in the
``Wigner-molecule'' regime of the charged fermions.

\textbf{\textit{Conclusions.}}
In summary, we calculated the persistent current in the strongly repulsive Hubbard model at finite temperatures from the
Bethe ansatz equations. The current shows several notable characteristics, including the temperature-dependent period
and the counterintuitive temperature dependence of the amplitude. To the best of our knowledge, this is the first example
of a temperature-dependent period of the persistent current, despite a large number of previous studies of
temperature-dependent PCs in many different models. It is quite unexpected, since the period is a fundamental quantum
property of the system which should be statistics-independent. An interesting future development would be
an extension of our results to any value of temperature by the alternative (but considerably more computationally
expensive) method of exact diagonalization and finding similar transport regimes in other, possibly
non-integrable, models of strongly interacting particles. We believe also that our findings will have considerable
implications for the interpretation of experiments on persistent currents of multi-component systems that can be
generated and investigated in the present-day atomtronics circuits \cite{RWMZ,AOC05,Amico21,Amico21b,PNLM,PNMA21,CPOCA}.

The work of O.I.P. has been supported  from the LAPLAS 6 program of the Romanian National Authority for Scientific Research (CNCS-UEFISCDI).
D.V.A. was supported by the US NSF grants EAGER 1836707 and 2104781.

\bigskip

\onecolumngrid
\newpage



\section*{Supplementary material}

# Supplemental Material: Temperature-dependent periodicity of the persistent current in strongly interacting systems

Ovidiu I. Pâţu[1] and Dmitri V. Averin[2]

[1]*Institute for Space Sciences, Bucharest-Măgurele, R 077125, Romania*
[2]*Department of Physics and Astronomy, Stony Brook University, Stony Brook, New York 11794, USA*

## I. STRONG COUPLING LIMIT OF THE BETHE ANSATZ EQUATIONS.

Strong coupling expansions of the BAEs for the Hubbard model can be found in [S1, S2] and Chap. III of [S3]. Introducing $u\Lambda_\alpha = \lambda_\alpha$ and taking the limit $u \to \infty$ in the BAEs of the Hubbard model threaded by magnetic flux

$$e^{ik_j L} = e^{2\pi i\phi/\phi_0} \prod_{\alpha=1}^{N_\downarrow} \frac{\lambda_\alpha - \sin k_j - iu}{\lambda_\alpha - \sin k_j + iu},$$

$$\prod_{j=1}^{N} \frac{\lambda_\alpha - \sin k_j - iu}{\lambda_\alpha - \sin k_j + iu} = \prod_{\beta=1,\beta\neq\alpha}^{N_\downarrow} \frac{\lambda_\alpha - \lambda_\beta - 2iu}{\lambda_\alpha - \lambda_\beta + 2iu},$$

with $j = 1, \cdots, N$ and $\alpha = 1, \cdots, N_\downarrow$ we obtain

$$e^{ik_j^\infty L} = e^{2\pi i\phi/\phi_0} \prod_{\alpha=1}^{N_\downarrow} \frac{\Lambda_\alpha - i}{\Lambda_\alpha + i}, \quad \text{(S-1)}$$

$$\left(\frac{\Lambda_\alpha - i}{\Lambda_\alpha + i}\right)^N = \prod_{\beta=1,\beta\neq\alpha}^{N_\downarrow} \frac{\Lambda_\alpha - \Lambda_\beta - 2i}{\Lambda_\alpha - \Lambda_\beta + 2i}, \quad \text{(S-2)}$$

which shows the decoupling of the charge and spin degrees of freedom. Eqs. (S-2) are the BAEs of the isotropic Heisenberg spin-chain with Hamiltonian

$$\mathcal{H}_{XXX} = -\frac{J}{4} \sum_{j=1}^{N} \left(\sigma_x^{(j)}\sigma_x^{(j+1)} + \sigma_y^{(j)}\sigma_y^{(j+1)}\right.$$
$$\left. + \sigma_z^{(j)}\sigma_z^{(j+1)} - \frac{1}{4}\right), \quad \text{(S-3)}$$

where $\sigma_{x,y,z}$ are the usual Pauli matrices and $J$ is the coupling strength. The energy and momentum of the spin chain are

$$E = \sum_{\alpha=1}^{N_\downarrow} \frac{2J}{\Lambda_\alpha^2 + 1}, \quad K = \left(\pi N_\downarrow - \sum_{\alpha=1}^{N_\downarrow} \theta(\Lambda_\alpha)\right) \mod 2\pi.$$

In logarithmic form, Eqs. (S-1) and (S-2) are

$$k_j^\infty L = 2\pi(I_j + \phi/\phi_0) + \sum_{\alpha=1}^{N_\downarrow} \theta(\Lambda_\alpha), \quad \text{(S-4)}$$

$$N\theta(\Lambda_\alpha) = 2\pi J_\alpha + \sum_{\beta=1,\beta\neq\alpha}^{N_\downarrow} \theta[(\Lambda_\alpha - \Lambda_\beta)/2]. \quad \text{(S-5)}$$

Using (S-5) in (S-4) and the antisymmetry of $\theta(x)$ we obtain the charge momenta in the impenetrable limit

$$k_j^\infty = \frac{2\pi}{L}\left(I_j + \frac{\phi}{\phi_0} + \sum_{\alpha=1}^{N_\downarrow} J_\alpha\right). \quad \text{(S-6)}$$

The first correction in $1/u$ is obtained by considering a general solution of the BAEs of the form (Chap. III of [S3])

$$k_j = k_j^\infty + \Delta k_j/u + O(1/u^2), \quad j = 1, \cdots, N, \quad \text{(S-7)}$$
$$\lambda_\alpha = u\Lambda_\alpha + O(1/u), \quad \alpha = 1, \cdots, N_\downarrow. \quad \text{(S-8)}$$

and equating terms of the same order. While $k_j^\infty$'s and $\Lambda_\alpha$'s are solutions of (S-1) and (S-2) the corrections satisfy the following system of equations

$$\Delta k_j L = 2 \sum_{\alpha=1}^{N_\downarrow} \frac{\Delta\Lambda_\alpha - \sin k_j^\infty}{\Lambda_\alpha^2 + 1}, \quad \text{(S-9)}$$

$$\sum_{j=1}^{N} \frac{\Delta\Lambda_\alpha - \sin k_j^\infty}{\Lambda_\alpha^2 + 1} = 2 \sum_{\beta=1}^{N_\downarrow} \frac{\Delta\Lambda_\alpha - \Delta\Lambda_\beta}{(\Lambda_\alpha - \Lambda_\beta)^2 + 4}. \quad \text{(S-10)}$$

Eq. (S-10) can be used in (S-9) to replace $\Lambda_\alpha$ obtaining

$$\Delta k_j = \frac{E_s}{L}\sum_{j=1}^{N}(\sin k_j^\infty - \sin k_n^\infty), \ E_s = -\frac{2}{N}\sum_{\alpha=1}^{N_\downarrow}\frac{1}{\Lambda_\alpha^2 + 1},$$

from which Eq. (3) of the main text can be derived.

When the parameter $\gamma = UL/N = U/n$ is large [S2], and for a given state of the spin-chain characterized by $\{\Lambda_\alpha\}_{\alpha=1}^{N_\downarrow}$, $\{J_\alpha\}_{\alpha=1}^{N_\downarrow}$, and a set of $\{I_j\}_{j=1}^{N}$, the charge momenta of a low-lying excitation can be expressed as

$$k_j = \frac{2\pi}{\tilde{L}}\left[I_j + \frac{\phi}{\phi_0} + \frac{1}{N}\sum_{\alpha=1}^{N_\downarrow} J_\alpha \right.$$
$$\left. - \frac{E_s}{Lu}\left(\sum_{j=1}^{N} I_j + N\frac{\phi}{\phi_0} + \sum_{\alpha=1}^{N_\downarrow} J_\alpha\right)\right], \quad \text{(S-11)}$$

with $\tilde{L} = L[1 - NE_s/(Lu)]$. From the definition of $\gamma = U/n$, we can see that Eq. (S-11) is valid in the case of very dilute systems $n \ll 1$ and moderate values of $U$, but also for any value of $0 < n < 1$ at very large values of the interaction strength $U$.

## II. GROUND STATE PROPERTIES OF THE HUBBARD MODEL

The ground state of the Hubbard model is described by the following system of integral equations [S4]:

$$\rho(k) = \frac{1}{2\pi} + \cos k \int_{-A}^{A} d\lambda \, a_1(\sin k - \lambda)\sigma(\lambda), \quad \text{(S-12a)}$$

$$\sigma(\lambda) = \int_{-Q}^{Q} dk \, a_1(\lambda - \sin k)\rho(k) - \int_{-A}^{A} d\lambda' \, a_2(\lambda - \lambda')\sigma(\lambda'), \quad \text{(S-12b)}$$

with the kernels defined by

$$a_l(x) = \frac{1}{2\pi}\frac{2lu}{(lu)^2 + x^2}, \quad u = \frac{U}{4t}. \quad \text{(S-13)}$$

For a system of $N$ electrons of which $N_\downarrow$ have spin down the parameters $Q$ and $A$ fix the particle density $n$ and magnetization per site $m$ via

$$n = \frac{N}{L} = \int_{-Q}^{Q} dk \, \rho(k), \quad \text{(S-14)}$$

$$m = \frac{N - 2M}{2L} = \frac{1}{2}\left[\int_{-Q}^{Q} dk \, \rho(k) - 2\int_{-A}^{A} d\lambda \, \sigma(\lambda)\right]. \quad \text{(S-15)}$$

The charge and spin velocities can be calculated from

$$v_c = \left.\frac{\bar{\kappa}'(k)}{2\pi\rho(k)}\right|_{k=Q}, \quad v_s = \left.\frac{\varepsilon'(k)}{2\pi\sigma(\lambda)}\right|_{\lambda=A}, \quad \text{(S-16)}$$

where $\bar{\kappa}'(k)$ and $\varepsilon'(k)$ are the derivatives of the dressed energies and satisfy

$$\bar{\kappa}'(k) = 2\sin k + \cos k \int_{-A}^{A} d\lambda \, a_1(\lambda - \sin k)\varepsilon'(\lambda),$$

$$\varepsilon'(\lambda) = \int_{-Q}^{Q} dk \, a_1(\lambda - \sin k)\bar{\kappa}'(k) - \int_{-A}^{A} d\lambda' a_2(\lambda - \lambda')\varepsilon'(\lambda').$$

The dressed charge matrix is defined by ([S5, S6, S7, S8] and Chap. VIII of [S3])

$$Z = \begin{pmatrix} Z_{cc} & Z_{cs} \\ Z_{sc} & Z_{ss} \end{pmatrix} = \begin{pmatrix} \xi_{cc}(Q) & \xi_{cs}(A) \\ \xi_{sc}(Q) & \xi_{ss}(A) \end{pmatrix}, \quad \text{(S-17)}$$

where $\xi_{ab}(k)$, $a, b \in \{c, s\}$ satisfy the system of integral equations

$$\xi_{cc}(k) = 1 + \int_{-A}^{A} d\lambda' \, \xi_{cs}(\lambda') a_1(\lambda' - \sin k), \quad \text{(S-18a)}$$

$$\xi_{cs}(\lambda) = \int_{-Q}^{Q} dk \, \cos k' \xi_{cc}(k') a_1(\sin k' - \lambda) - \int_{-A}^{A} d\lambda' \, \xi_{cs}(\lambda') a_2(\lambda' - \lambda), \quad \text{(S-18b)}$$

$$\xi_{sc}(k) = \int_{-A}^{A} d\lambda' \xi_{ss}(\lambda') a_1(\lambda' - \sin k), \quad \text{(S-18c)}$$

$$\xi_{ss}(\lambda) = 1 + \int_{-Q}^{Q} dk' \cos k' \xi_{sc}(k') a_1(\sin k' - \lambda) - \int_{-A}^{A} d\lambda' \xi_{ss}(\lambda') a_2(\lambda' - \lambda). \quad \text{(S-18d)}$$



\end{document}